\documentclass[review]{article}  %elsarticle
\usepackage[utf8]{inputenc}
\usepackage[english]{babel}
\usepackage[T1]{fontenc}

\usepackage[a4paper,top=2.5cm,bottom=2.5cm,left=2.5cm,right=2.5cm,marginparwidth=2cm]{geometry} 
\usepackage{setspace}
\usepackage{float}
\usepackage{relsize}
\usepackage{lipsum}

\usepackage{amsmath}
\usepackage{amssymb}
\usepackage{bm}
\usepackage{bbm}
\usepackage{outlines}  %nested bullet point lists
\usepackage{graphicx}
\usepackage{rotating}
\usepackage{authblk}
\usepackage[round]{natbib}
\usepackage{verbatim}
\usepackage{url}
\usepackage{algorithm}
\usepackage{algpseudocode}
\usepackage{multirow}
\usepackage[colorinlistoftodos]{todonotes}   %disable
\usepackage[mathlines, displaymath]{lineno}
\usepackage{ulem} % for strikethrough: \sout{text}
\usepackage{cleveref}   %always the last one

\graphicspath{{Figures}}

\usepackage{etoolbox}
\patchcmd{\pprintMaketitle}%
{\vskip36pt} % original vertical space 36 pt
{\vskip10pt} % minimum vertical space  <<<<<<<<<<<
{}
{}
\makeatletter
\let\LN@align\align
\let\LN@endalign\endalign
\renewcommand{\align}{\linenomath\LN@align}
\renewcommand{\endalign}{\LN@endalign\endlinenomath}
\let\LN@gather\gather
\let\LN@endgather\endgather
\renewcommand{\gather}{\linenomath\LN@gather}
\renewcommand{\endgather}{\LN@endgather\endlinenomath}
\makeatother

\makeatletter
\def\ps@pprintTitle{%
 \let\@oddhead\@empty
 \let\@evenhead\@empty
 \def\@oddfoot{\centerline{\thepage}}%
 \let\@evenfoot\@oddfoot}
\makeatother

\makeatletter
\def\ps@pprintTitle{%
 \let\@oddhead\@empty
 \let\@evenhead\@empty
 \def\@oddfoot{\centerline{\thepage}}%
 \let\@evenfoot\@oddfoot}
\makeatother

\newsavebox\extrainfobox

\makeatletter
\renewenvironment{abstract}
 {\small
  \begin{center}
  \bfseries \abstractname\vspace{-0.5em}\vspace{0pt}
  \end{center}
  \list{}{
    \setlength{\leftmargin}{0cm}%
    \setlength{\rightmargin}{\leftmargin}%
  }%
  \item\relax}
 {\endlist}

\title{A Bayesian location-scale joint model for time-to-event and multivariate longitudinal data with association based on within-individual variability}

\date{\today}

\author[1]{Marco Palma\thanks{Corresponding author: marco.palma@mrc-bsu.cam.ac.uk}}
\author[2]{Ruth H Keogh}
\author[3a,3b]{Siobhán B Carr}
\author[4a,4b]{Rhonda Szczesniak}
\author[5]{David Taylor-Robinson}
\author[6]{Angela M Wood}
\author[7a,7b]{Graciela Muniz-Terrera%\footnote{Joint senior authors}
}
\author[1]{Jessica K Barrett} %the footnote applies both to Muniz-Terrera and Barrett

\affil[1]{MRC Biostatistics Unit, University of Cambridge, UK}
\affil[2]{Department of Medical Statistics, London School of Hygiene \& Tropical Medicine, UK}
\affil[3a]{Royal Brompton Hospital, part of Guy's and St Thomas' NHS Foundation Trust, UK}
\affil[3b]{National Heart and Lung Institute, Imperial College, London, UK}
\affil[4a]{Divisions of Biostatistics and Epidemiology and Pulmonary Medicine, Cincinnati Children's Hospital Medical Center, United States}
\affil[4b]{Department of Pediatrics, University of Cincinnati, United States}
\affil[5]{Department of Public Health, Policy and Systems, University of Liverpool, UK}
\affil[6]{Cardiovascular Epidemiology Unit, University of Cambridge, UK}
\affil[7a]{Ohio University Heritage College of Osteopathic Medicine, United States}
\affil[7b]{University of Edinburgh, UK}

\begin{document}

\maketitle

%\pagenumbering{gobble}
%\singlespacing
\doublespacing

%\begin{frontmatter}

\begin{abstract}
Within-individual variability of health indicators measured over time is becoming commonly used to inform about disease progression. Simple summary statistics (e.g.\ the standard deviation for each individual) are often used but they are not suited to account for time changes. In addition, when these summary statistics are used as covariates in a regression model for time-to-event outcomes, the estimates of the hazard ratios are subject to regression dilution. To overcome these issues, a joint model is built where the association between the time-to-event outcome and multivariate longitudinal markers is specified in terms of the within-individual variability of the latter. A mixed-effect location-scale model is used to analyse the longitudinal biomarkers, their within-individual variability and their correlation. The time to event is modelled using a proportional hazard regression model, with a flexible specification of the baseline hazard, and the information from the longitudinal biomarkers is shared as a function of the random effects. The model can be used to quantify within-individual variability for the longitudinal markers and their association with the time-to-event outcome. We show through a simulation study the performance of the model in comparison with the standard joint model with constant variance. The model is applied on a dataset of adult women from the UK cystic fibrosis registry, to evaluate the association between lung function, malnutrition and mortality.
\end{abstract}

%\begin{keyword}

%\end{frontmatter}
\newpage

%\begin{linenumbers}

\section{Introduction}

Longitudinal data are ubiquitous in biomedical studies, where often the evolution of a biomarker over time is monitored to evaluate health status or disease progression. The focus of the analysis is on the trend over time, where steady or sudden changes might be indicative of changes in the health status of an individual. The information about the trend can also be used within a joint model for longitudinal and time-to-event data, where the predicted value for a longitudinal biomarker is used to assess risk of death or another outcome \citep{tsiatis2004joint, rizopoulos2012joint, diggle2024longitudinal}.

A push towards a richer characterisation of longitudinal data beyond the trend has come from different directions. In particular, the notion of visit-to-visit (or long-term) within-individual variability (WIV), used to quantify fluctuations around the mean, has gained interest in order to answer clinical questions about chronic disease and biomarker monitoring. % In cardiovascular research, \citet{rothwell2010prognostic} has first suggested that blood pressure variability should be taken into account (in addition to the mean) as predictor for stroke, highlighting potential need for blood pressure-stabilising drugs. A later review and meta-analysis by \citet{stevens2016blood} has collected studies reporting about long-term and short-term variability in blood pressure and found association with many cardiovascular outcomes, as well as mortality.
First considered in cardiovascular research (\citealp{rothwell2010prognostic} first suggested that blood pressure variability should be taken into account as predictor for stroke), modelling WIV has proved particularly popular in cystic fibrosis (CF) studies. Lung function WIV is a key quantity of interest in CF, as it is often used as an alternative measure to the number of pulmonary exacerbations (acute events associated with decline in CF) whose definition is not fully standardised \citep{stanford2021pulmonary}. For example, \citet{morgan2016forced} argued that lung function variability is predictive of future lung function decline, while \citet{todd2024forced} used a cumulative measure of lung function variability within a Cox proportional hazard model for children and adults with CF. \citet{szczesniak2025real} evaluated changes in lung function variability linked to the initiation of a new CF treatment.

In these studies, within-individual variability is quantified using summary statistics (e.g.\ standard deviation of the individual observations) or differences with respect to a reference level (%that could be 
the median or the maximum value, possibly within a certain time-window). These two-stage approaches, although computationally appealing and immediately interpretable, may incur in biases (e.g.\ immortal time, when the computation is based on a minimum number of observations) and their precision is rather sensitive to the number of visits (which is often low in biomedical data). In addition, \citet{barrett2019estimating} showed that when such measures of variability are used in a time-to-event model, their association with the outcome is not accurately estimated and suffers from regression dilution bias towards zero.

These limitations motivated novel methodological research in joint modelling to properly address the problem of WIV quantification. \citet{gao2011joint} and \citet{barrett2019estimating} introduced a Bayesian joint model where a Bayesian mixed-effects location-scale model (MELSM, \citealp{hedeker2008application, hedeker2012modeling}) is fitted for the longitudinal biomarker, with a Markov chain Monte Carlo scheme. In this model, both the mean and the residual variability can be expressed as functions of time, covariates as well as subject-specific random effects, and both the current values of mean and variability are included in the time-to-event submodel. In a similar fashion, \citet{courcoul2023location} and \citet{li2023joint} proposed frequentist approaches to fit a joint model where the variability of a single biomarker is associated to the hazard of two competing events (of which one is mortality). \citet{courcoul2024joint} extended this setting for the case of multiple measurements for the same biomarker taken at the same visit, modelling intra-visit variability as well as WIV. 

Nevertheless, while multivariate longitudinal models have been explored, even in a location-scale setting \citep{kapur2015bayesian}, few attempts have been made so far at modelling WIV where (potentially correlated) multiple longitudinal biomarkers are jointly considered with a time-to-event outcome. The Bayesian additive model for location, scale and shape (bamlss, \citealp{umlauf2018bamlss}) offers several approaches to estimate WIV within a joint model (\citealp{kohler2017flexible, kohler2018nonlinear}). More recently \citet{volkmann2023flexible}, with a more parsimonious representation of the mean using functional principal component analysis, provided another approach to model concurrently both the mean and the variability. Nevertheless, none of these models permit the inclusion of the WIV association term within the time-to-event submodel.

We propose therefore a Bayesian multivariate joint model with WIV-based association (JM-WIV). The longitudinal submodel accounts for modelling both the mean and the within-individual variability of more than one biomarker, as well as the correlation between biomarkers, using a shared random effect structure. The time-to-event submodel is a proportional hazard model that includes a function of the variability in the longitudinal biomarkers, with a flexible specification of the baseline hazard based on B-splines. We provide a novel implementation in Stan for a bivariate longitudinal setting. 

The paper is structured as follows. \Cref{sec:methods} illustrates the structure of the model, with a description of the model assumptions and the key aspects of model fitting. A simulation study showing the advantage of this model against the standard joint model in the literature is provided in \Cref{sec:simulation}. Next, in \Cref{sec:appl} we apply the model on a dataset of women from the UK Cystic Fibrosis registry, where we evaluate the relationship between lung function, body mass index and mortality. Lastly, we discuss the findings, limitations and further developments in \Cref{sec:discussion}.

\section{Methods}\label{sec:methods}

\subsection{JM-WIV model formulation}

\begin{comment}
\begin{outline}
\1 Longitudinal submodel: MELSM \citep{hedeker2008application}
    \2 Location and scale components
\1 Time-to-event submodel: proportional hazards regression 
\1 Association structures and their functional forms
    \2 Random effects for both location and scale
    \2 Current (estimated) values of mean (and WIV?)
    \2 All other association structures already available?
\1 Assumptions
    \2 Full Conditional Independence - shared random effects 
    \2 uninformative censoring
\end{outline}

\newpage
\end{comment}

\subsubsection{Longitudinal submodel}

Consider $K$ longitudinal biomarkers $y_{ijk}$ ($k = 1,\dots, K$) for the $i$-th subject ($i = 1,\dots, N$) and $j$-th occasion ($j = 1,\dots, J_i$), observed at subject-specific ordered time points $\bm{t_i} = \left[t_{i1}, \dots, t_{iJ_i}\right]^\mathtt{T}$, with $t_{i1} \leq \dots \leq t_{iJ_i}$. As opposed to standard linear mixed models, the independent error terms are assumed normally distributed with non-constant variance: $\varepsilon_{ijk} \sim N(0, \sigma^2_{ijk})$. A mixed-effect location-scale model \citep{hedeker2008application} is proposed for the outcome:
\begin{align}
    y_{ijk} &= \eta^{\mu_k}_{i}(t_{ij}) + \varepsilon_{ijk} \notag\\
    g(\sigma_{ijk}) &= \eta^{\sigma_k}_{i}(t_{ij})
\end{align}
where $g(\cdot) = log(\cdot)$ to ensure that the variance is non-negative. Note that the formulation for the standard deviation could be also rewritten in terms of the variance and viceversa. For $\psi \in \{\mu, \sigma\}$%$\psi \in \{\mu_1, \dots, \mu_K, \sigma_1, \dots, \sigma_K\}$
, the fixed-effect design matrix $X_{ij}^{\psi_k}$ and the random-effect design matrix $Z_{ij}^{\psi_k}$ in the structural additive predictor $\eta^{\psi_k}(t)$ can potentially include baseline information or (the basis functions for the additive modelling of) time-varying covariates:
\begin{equation*}
    \eta^{\psi_k}_{i}(t_{ij}) = X_{ij}^{\psi_k}\bm{\beta}^{\psi_k} + Z_{ij}^{\psi_k} \bm{b}_{i}^{\psi_k} %+ \sum_{m=1}^{M_\psi} s_m^\psi(\bm{\tilde{x}}_m(t), t) 
\end{equation*}
The components of the longitudinal submodel are linked via the random effect vector $\bm{b}_i = \left[\bm{b}^{\bm{\mu}}_i, \bm{b}^{\bm{\sigma}}_i\right]^\mathtt{T} \sim N(\bm{0}, \Sigma)$.

We consider an unstructured covariance matrix $\Sigma = TPT$ with $T = \text{diag}(T^{\mu_1}, \dots, T^{\mu_K}, T^{\sigma_1}, \dots, T^{\sigma_K})$. The block $T^{\psi_k}$ is diagonal with elements $\tau^{\psi_k}_l$ ($l = 1,\dots, L^{\psi_k}$) corresponding to the standard deviation of the $l$-th element of the random effect vector $\bm{b}^{\psi_k}$. The correlation matrix $P$ takes the form 
%
\begin{comment}   
\begin{equation}
    P = 
    \begin{bmatrix}
    P^{\mu_1\mu_1}& \cdots & P^{\mu_1\mu_K} & P^{\mu_1\sigma_1} & \cdots & P^{\mu_1\sigma_K} \\    
     & \ddots &  & & & \\ 
    P^{\mu_K\mu_1} & & P^{\mu_K\mu_K} & & & \\ 
    P^{\sigma_1\mu_1} & & & P^{\sigma_1\sigma_1} & & \\ 
    & & & & \ddots & \\ 
    P^{\sigma_K\mu_1}& & & & & P^{\sigma_K\sigma_K}\\ 
    \end{bmatrix}
\end{equation}
\end{comment}
%
\begin{equation*}
    P = 
    \begin{bmatrix}
    P^{\mu_1\mu_1}& P^{\mu_1\sigma_1} & P^{\mu_2 \sigma_1} &\cdots &  & P^{\mu_1\sigma_K} \\    
    P^{\sigma_1\mu_1}  & P^{\sigma_1\sigma_1}  &  & & & \\ 
    P^{\mu_2\mu_1} & & P^{\mu_2\mu_2} & & & \\ 
    %P^{\sigma_1\mu_1} & & & \ddots & & \\ 
    \vdots &  & &  \ddots & \\ 
    P^{\sigma_K\mu_1}& & & & & P^{\sigma_K\sigma_K}\\ 
    \end{bmatrix}
\end{equation*}
with $P^{{\psi_k}{\psi'_{k'}}} = (P^{{\psi'_{k'}}{\psi_k}})^\mathtt{T}$. For example, $P^{{\mu_1}{\mu_2}}$ refers to the correlation matrix for the random effects of the means of biomarker 1 and 2, and $P^{{\mu_1}{\sigma_1}}$ refers to the correlation matrix for the random effects of the mean and the standard deviation of biomarker 1. These matrices reduce to a single number if only random intercepts are used.
%
%
%
\begin{comment}
The matrix $P^{{\psi_k}{\psi'_{k'}}}$ has elements 
%
\begin{equation}
    P_{ll'}^{{\psi_k}{\psi'_{k'}}} = 
    \begin{cases}
        1 & \text{if }\psi_k = \psi'_{k'} \land l = l'\\
        \rho_{ll'}^{\psi\psi'} & \text{otherwise}
    \end{cases}
\end{equation}
%
\end{comment}
%
%
%
For the matrix $P^{{\psi_k}{\psi'_{k'}}}$, the element $\rho_{l,l'}^{{\psi_k}{\psi'_{k'}}}$ denotes the correlation between the random effects $b_l^{\psi_k}$ and $b_{l'}^{\psi'_{k'}}$ ($l' = 1,\dots, L^{\psi'_{k'}}$). It is equal to 1 if $\psi_{k} = \psi'_{k'} \land l = l'$ (i.e.\ the diagonal elements are all equal to 1).

\subsubsection{Event submodel}
Let $T_i$ be the observed right-censored time ($T_i \geq t_{iJ_i}$) %, $T_i = \min{T_i^*, C_i}$, $T_i^*$ being the real event time and $C_i$ the censoring event 
and $\delta_i$ the event indicator taking value 1 if the event occurred for the $i$-th subject and 0 otherwise. The event submodel is specified as a proportional hazard model of the form
\begin{equation}\label{eq:eventsubmodel}
    h_i(t|\mathcal{L}_i(t)) = \exp\{\log h_0(t) + \eta^\gamma_i + \eta^\alpha_i(\mathcal{L}_i(t);t)\}
\end{equation}
where $\mathcal{L}_i(t)$ incorporates the information about the longitudinal history, $\eta^\gamma_i$ is a linear predictor for the baseline covariates of the time-to-event submodel, and $\eta^\alpha_i(\mathcal{L}_i(t);t)$ accounts for the association between the hazard function and the longitudinal biomarkers.

%\subsubsection{Association structure}
Let us define the current value of the linear predictor for the mean and standard deviation of all longitudinal biomarkers $\bm{\eta}_i(t) = \left[\eta^{\mu_1}_i(t),\dots,\eta^{\mu_K}_i(t), \eta^{\sigma_1}_i(t), \dots, \eta^{\sigma_K}_i(t)\right]$. The association between the longitudinal and time-to-event submodels can be specified as:
\begin{outline}
    \1 a function of the current value (CV) of the mean and standard deviation $$\eta^{\alpha_{CV}}_i(\mathcal{L}_i(t);t) = f(\bm{\eta}^\mu_i(t), \exp(\bm{\eta}^\sigma_i(t)))$$
    \1 a function of the current linear predictor of the mean and standard deviation (LP) $$\eta^{\alpha_{LP}}_i(\mathcal{L}_i(t);t) = f(\bm{\eta}^\mu_i(t), \bm{\eta}^\sigma_i(t))$$
    \1 a function of the random effects (RE) for the mean and standard deviation
    $$\eta^{\alpha_{RE}}_i(\mathcal{L}_i(t);t) = f(\bm{b}_i)$$
\end{outline}
Contrary to standard joint models, all these association structures incorporate now information about the within-individual variability. Other forms of association (as function of lagged predicted value, slope, area under the curve of the linear predictors and interactions, like in \citealt{kohler2018nonlinear}) can also be specified for both parameters. 

\subsubsection{Assumptions}
As in standard joint model \citep{rizopoulos2012joint}, we assume that the $N$ subjects are independent from each other and the right-censoring is uninformative. Furthermore, conditional on the random effects,
\begin{outline}
\1 there is independence between the longitudinal outcomes and between each longitudinal outcome and the time-to-event one;
%
%\begin{equation}
%    p(\bm{y}_{ik}, T_i, \delta_i | \bm{b}_i) = p(T_i, \delta_i | \bm{b}_i)\prod_k p(y_{ik}| \bm{b}_i)
%\end{equation}
%
\1 the repeated measurements are independent of each other.
%
%\begin{equation}
%    p(\bm{y}_{ik}| \bm{b}_i) = \prod_j p(y_{ijk}| \bm{b}_i)
%\end{equation}
%
\end{outline}
The combined effect of the assumptions above is the following decomposition of the likelihood function:
\begin{equation}\label{eq:decomp_assumptions}
    p(y_{ijk}, T_i, \delta_i | \bm{b}_i) = p(T_i, \delta_i | \bm{b}_i)\prod_{k=1}^K \prod_{j=1}^{J_i} p(y_{ijk}| \bm{b}_i)
\end{equation}

\subsection{Model estimation}

%\begin{outline}
%\1 Formula of the log-posterior distribution
%    \2 Quadrature for hazard function
%    \2 More flexible ways available in Stan
%\end{outline}

%\subsubsection{Likelihood function}

Let $\bm{\theta}$ be the fixed parameter vector. By the assumptions in \Cref{eq:decomp_assumptions}, we can factorise the likelihood function into event and longitudinal processes.
\begin{align}
    p(\bm{T}, \bm{\delta}, \bm{y} | \bm{\theta}, \bm{b}) & = \prod_{i=1}^N p( T_i, \delta_i, \bm{y}_i | \bm{\theta}, \bm{b}_i)\notag\\
    & = \prod_{i=1}^N p^{event}(T_i, \delta_i|\bm{b_i}, \bm{\theta})p^{long}(\bm{y_i}|\bm{b_i}, \bm{\theta}) \notag\\
    & = \prod_{i=1}^N p^{event}(T_i, \delta_i|\bm{b_i}, \bm{\theta})\prod_{k=1}^K \prod_{j=1}^{J_i} p^{long}(y_{ijk}| \bm{b}_i, \bm{\theta})\notag
\end{align}
The log-likelihood is therefore equal to 
\begin{align}
    \log\left(p(\bm{T}, \bm{\delta}, \bm{y} | \bm{\theta}, \bm{b})\right) &= \sum_{i=1}^N \left[ \log(p^{event}(T_i, \delta_i|\bm{b_i}, \bm{\theta})) + \sum_{k=1}^K \sum_{j=1}^{J_i}  \log(p^{long}(y_{ijk}|\bm{b_i}, \bm{\theta})) \right].\notag
\end{align}
For the longitudinal submodel, we use the normal log-likelihood of the form
\begin{align}
    \log(p^{long}(y_{ijk}|\bm{b_i}, \bm{\theta})) & = -\dfrac{1}{2}\log(2\pi) - \log(\sigma_{ijk}) - \dfrac{1}{2\sigma_{ijk}^2}\left(y_{ijk} - \mu_{ijk}\right)^2 \notag\\
    & = - \dfrac{1}{2}\log(2\pi) - \eta^{\sigma_k}_{i}(t_{ij}) - \dfrac{1}{2\exp(2\eta^{\sigma_k}_{i}(t_{ij}))}\left(y_{ijk} - \eta^{\mu_k}_{i}(t_{ij})\right)^2
\end{align}
where $\sigma_{ijk}$ is replaced by the corresponding function of the linear predictor for the variability submodel.
For the proportional hazard time-to-event submodel, the standard likelihood formulation is used \citep{kalbfleisch2011statistical}
\begin{equation}\label{eq:loglik_event}
    \log(p^{event}(T_i, \delta_i|\bm{b_i}, \bm{\theta})) = \delta_i\log{\left(h_i(T_i|\mathcal{L}_i(T_i))\right)} - \int_0^{T_i} h_i(s|\mathcal{L}_i(s)) ds %\\     & = h_i(T_i|\mathcal{L}_i(T_i))^{\delta_i} \exp\left(- \int_0^{T_i} h_i(s|\mathcal{L}_i(s))) ds\right)\\
    %& =     h_i(t|\mathcal{L}_i(t)) = \exp\{\log h_0(t) + \eta^\gamma_i + \eta^\alpha_i(\mathcal{L}_i(t);t)\}
\end{equation}
where $h_i(\cdot)$ is given in \Cref{eq:eventsubmodel}. The logarithm of $h_0(t)$ (the baseline hazard) is modelled using B-splines. The integral in \Cref{eq:loglik_event} is approximated using Gauss--Kronrod quadrature with $Q$ nodes, weights $w_q$ and locations $s_q\;(q= 1,\dots, Q)$:
\begin{equation}
    \int_0^{T_i} h_i(s|\mathcal{L}_i(s)) \approx \dfrac{T_i}{2}\sum_{q=1}^Qw_qh_i\left(\dfrac{T_i(1+s_q)}{2}\right).
\end{equation}
We use common priors for the coefficients. Normally distributed priors are used for the fixed effects for the longitudinal submodel and the baseline covariates for the time-to-event submodel. For the random effects, the standard deviations are drawn from a half-t distribution, while the LKJ prior is used for the correlation matrix. A normal distribution is used as prior for the B-spline coefficients of the log baseline hazard.

\subsubsection{Implementation in Stan}

Our implementation of the multivariate JM-WIV (available at \url{https://github.com/marcopalma3/rstanjmwiv}) extends the joint model in Stan by \citet{brilleman2018StanCon}, now included in the \verb|R| package \verb|rstanarm| \citep{rstanarm}. Estimation of the model is performed via Hamiltonian Monte Carlo, which explores the parameter space more efficiently than MCMC methods (used e.g.\ in bamlss \citealp{kohler2017flexible, volkmann2023flexible}). The current version of the software handles a bivariate longitudinal setting and allows for three association structures: RE, LP and CV. The user can specify the hyperparameters of prior distributions, the number $Q$ of quadrature points (the default used in this work is $Q = 15$) and the number of basis functions for the log baseline hazard functions (here we consider 6 cubic splines). The code and the results in \Cref{sec:simulation} and \Cref{sec:appl} are available on \url{https://github.com/marcopalma3/JM-WIV}.  

\section{Simulation study}
\label{sec:simulation}

We illustrate the benefit of our modelling approach via a simulation study, with two aims. First, we show that the inference procedure returns results in line with the simulated data. Second, we compare inference performances with standard joint models (as implemented in the R package JMbayes2) which do not offer the possibility of modelling the association of within-individual variability with the event of interest.

We designed a simulation scenario where the correct model includes 
association between the WIV of the longitudinal biomarkers and time-to-event submodel, adapting the structure illustrated in \citet{alvares2021tractable} to the MELSM case. We consider the case of $K=2$ normally-distributed longitudinal biomarkers with both mean and standard deviation dependent on a fixed time effect and a random intercept:
\begin{align}\label{eq:SIM_MELSM}
    %y_{ijk} &= \beta_0^{\mu_k} + \beta_1^{\mu_k}t_{ij} + b^{\mu_k}_{i} + \varepsilon_{ijk} \notag\\
    \mu_{ijk} &= \beta_0^{\mu_k} + \beta_1^{\mu_k}t_{ij} + b^{\mu_k}_{i} \notag\\
    \log(\sigma_{ijk}) &= \beta_0^{\sigma_k} + \beta_1^{\sigma_k}t_{ij} + b^{\sigma_k}_{i}
\end{align}
To generate the longitudinal observations, we consider a regular schedule every 1 time unit from 0 to $T_i$ (or $T_i^*$ for censored observations). 

For the time-to-event submodel, we specify a Cox model with constant baseline hazard function (also known as Cox-Exponential model, \citealp{bender2005generating}) and two baseline covariates ($w_1$ binary, $w_2$ normally distributed). The association parameters for each biomarker are based on the current value of the linear predictors (LP) $\mu_{ijk}(t)$ and $\log{(\sigma_{ijk}(t))}$: 
\begin{equation}\label{eq:SIM_eventsubmodel}
    h_i(t|\mathcal{L}_i(t)) = \exp\{\log \lambda + w_{1i}\gamma_1 + w_{2i}\gamma_2 + \sum_{k=1}^2 \left[\alpha^{\mu_k}\mu_{ik}(t) + \alpha^{\sigma_k}\log{(\sigma_{ik}(t))} \right]   
    \}
\end{equation}
%
%First, we generate random effects from a multivariate normal distribution with parameters <ADD PARAMETERS>. Then we simulate time-to-event times under a Cox-exponential specification with parameter $\lambda=1$ and $U_i \sim Unif(0,1)$ following the approach illustrated in \citep{bender2005generating}
%
%\begin{equation}
%       T_i = - \dfrac{log(U_i)}{\lambda\exp(w_i^\mathtt{T}\gamma + \bm{b}_i^\mathtt{T}\alpha)}.
%\end{equation}
%
%We specify the censoring probability to be equal to $p^{cens} = 0.2$ and for the censored observations we replace $T_i$ by $T_i^* \sim Unif (0, T^{max})$ %where $T^{max} = <ADD VALUE>$ is the length of the study
%
\begin{algorithm}
\caption{Simulation from Cox-Exponential JM-WIV model with LP association}\label{alg:cap}
\begin{algorithmic}
\State For $i = 1, \dots, N$ individuals
\State \textbf{Random effects.} Simulate $\bm{b}_i \sim N_4(0, TPT)$
\State \textbf{Time-to-event process.} 
\State\hspace{\algorithmicindent} Draw $w_{1i}$ (from a binomial distribution) and $w_{2i}$ (from a standard normal distribution)
\State\hspace{\algorithmicindent} Specify $A$ and $B$ as follows:
    \begin{align*}
       A &= \sum_{k=1}^2 \left[\alpha^{\mu_k}(\beta_0^{\mu_k} +  b^{\mu_k}_{0i}) + \alpha^{\sigma_k}(\beta_0^{\sigma_k} +  b^{\sigma_k}_{0i}) \right]  \\
       B &= \alpha^{\mu_k}\beta_1^{\mu_k} + \alpha^{\sigma_k}\beta_1^{\sigma_k}
    \end{align*}
\hspace{\algorithmicindent} so that 
    \begin{equation*}
          h_i(t) = \exp\{\log \lambda + \bm{w}_i^\mathtt{T}\bm{\gamma} + A + Bt\}
    \end{equation*}
\State\hspace{\algorithmicindent} Draw $U_i \sim Unif(0,1)$ and generate event times:
\begin{equation*}
    T_i = \dfrac{1}{B}\log\left[1 - \dfrac{B \log(U_i)}{\lambda\exp\left(\bm{w}_i^\mathtt{T}\bm{\gamma} + A\right)}\right]
\end{equation*}
\State\hspace{\algorithmicindent} Generate independent censoring time $T_i^* \sim Unif(0, T^{max})$ 
\State\hspace{\algorithmicindent} Assign event indicator $\delta_i = 1$ if $T_i \leq T_i^*$ (zero otherwise)
\State \textbf{Longitudinal process.}
\State\hspace{\algorithmicindent} Generate measurement schedule at regular times between 0 and $T^{max}$.
\State\hspace{\algorithmicindent} Generate the biomarker values following the MELSM model in \Cref{eq:SIM_MELSM}.
\end{algorithmic}
\end{algorithm}
%
%
%
%The longitudinal outcomes $y_{ij1}$ and $y_{ij2}$ follow a MELSM specification with a fixed effect and a random intercept for both the mean and the logarithm of the standard deviation:
%
%\begin{equation}
%    \eta^{\psi_k}_{i}(t_{ij}) = \bm{\beta}_0^{\psi_k} + year_{ij}^{\psi_k}\bm{\beta}_1^{\psi_k} + b_{i}^{\psi_k}.
%\end{equation}
%
The parameters of the simulation study are reported in \Cref{tab:simpar} and in the Supplementary Material. For the longitudinal submodel, the correlation structure for the random effects is approximately equal to the corresponding variance-covariance matrix in the CF registry data application in \Cref{sec:appl}. In particular, there is positive correlation between mean and standard deviation in both longitudinal biomarkers, and a positive correlation between the means of the two outcomes.

We simulate 200 datasets with $N=1000$ individuals each. We compare our model (JM-WIV) with the standard joint model as implemented in the \verb|R| package JMbayes2 \citep{JMbayes2}, both with the correct mean specification for the longitudinal submodels (while the standard deviation in JMbayes2 is assumed constant). %We compute bias and coverage for eight coefficients: the baseline covariate effects $\gamma_1$  and $\gamma_2$ for the time-to-event submodel, the association parameters $\alpha^{\mu_1}$ and $\alpha^{\mu_2}$ between the mean of the two biomarkers and the hazard, and the two intercepts and two fixed time effects ($\bm{\beta}$) for the longitudinal submodel. 
The joint model in JMbayes2 is fitted with a burn-in of 1000 iterations and a total number of 11000 iterations. The number of iterations for JM-WIV is 6000, of which 3000 used as burn-in. The results for bias and coverage are reported in \Cref{tab:simpar}.
%
\begin{comment}
    
\begin{table}[hbtp]
\centering
\begin{tabular}{l|r|rrr|rrr}
%\hline
    \multirow{2}{*}{\bfseries} & \multirow{2}{*}{\bfseries Input} &\multicolumn{3}{|c|}{\bfseries JM-WIV}  & \multicolumn{3}{c}{\bfseries JMbayes2}\\
%    \cline{1-8}
%    {\bfseries } & {\bfseries OBR} &{\bfseries AR}  & {\bfseries OBR} &{\bfseries AR} & {\bfseries OBR} &{\bfseries AR}\\
    %\hline

  &  & Mean & eSD & Coverage & Mean & eSD & Coverage\\
\hline
%\hline
$\gamma_1$ & 0.930 & 0.908 & 0.122 & 0.960 & 0.892 & 0.129 & 0.970\\
%\hline
$\gamma_2$ & -2.300 & -2.264 & 0.110 & 0.940 & -2.197 & 0.106 & 0.805\\
%\hline
$\alpha^{\mu_1}$ & -2.240 & -2.153 & 0.138 & 0.940 & -2.097 & 0.149 & 0.775\\
%\hline
$\alpha^{\mu_2}$ & 0.550 & 0.529 & 0.184 & 0.995 & 0.592 & 0.241 & 0.890\\
\hline
%\hline
$\beta^{\mu_1}_0$ & 2.190 & 2.189 & 0.029 & 0.915 & 2.192 & 0.029 & 0.910\\
%\hline
$\beta^{\mu_1}_1$ & -0.040 & -0.040 & 0.002 & 0.935 & -0.040 & 0.003 & 0.930\\
%\hline
$\beta^{\mu_2}_0$ & 1.040 & 1.040 & 0.032 & 0.965 & 1.042 & 0.033 & 0.960\\
%\hline
$\beta^{\mu_2}_1$ & 0.350 & 0.351 & 0.009 & 0.950 & 0.351 & 0.010 & 0.955\\
\hline
\end{tabular}

%\caption{Simulation results. For each model, the mean and standard devation (eSD) of the 200 posterior means are reported, alongside the coverage probability (proportion of times the input value is included in the 95\% credible interval).}
\label{tab:simpar}
\end{table}
\end{comment}

The simulation shows that the estimation of the parameters of the means of the longitudinal biomarkers remains essentially unaffected by the modelling of the standard deviation. In the survival submodel, we observe a difference between the two models: the mean estimates in JMbayes are on average further from the simulated value compared to JM-WIV, and the coverage is well below the nominal level of 0.95 (except for $\gamma_1$). The mean estimates for JM-WIV (i.e.\ the correctly specified model in this instance) are slightly shrunk towards zero but the coverage is consistent with the nominal level, while the association with WIV is under coverage only for the first longitudinal outcome.  
\begin{table}[hbtp]
    \centering
\begin{tabular}{l|r|rrr|rrr}
%\hline
    \multirow{2}{*}{\bfseries} & \multirow{2}{*}{\bfseries Input} &\multicolumn{3}{|c|}{\bfseries JM-WIV}  & \multicolumn{3}{c}{\bfseries JMbayes2}\\
  &  & Mean & eSD & Coverage & Mean & eSD & Coverage\\
\hline
\begin{comment}
    
\hline
\hline
y1..Intercept. & 2.19 & 2.187 & 0.023 & 0.975 & 2.189 & 0.024 & 0.970\\
\hline
y1.visit\_times & -0.04 & -0.040 & 0.002 & 0.945 & -0.040 & 0.002 & 0.955\\
\hline
b\_sd1 & 0.81 & 0.812 & 0.019 & 0.950 & - & - & -\\
\hline
y1sigma..Intercept. & -1.36 & -1.367 & 0.028 & 0.970 & -1.239 & 0.030 & 0.000\\
\hline
y1sigma.visit\_times & -0.02 & -0.020 & 0.007 & 0.960 & - & - & -\\
\hline
b\_sd2 & 0.44 & 0.442 & 0.019 & 0.960 & - & - & -\\
\hline
y2..Intercept. & 1.04 & 1.042 & 0.033 & 0.960 & 1.044 & 0.033 & 0.935\\
\hline
y2.visit\_times & 0.35 & 0.350 & 0.009 & 0.970 & 0.351 & 0.009 & 0.950\\
\hline
b\_sd3 & 0.52 & 0.517 & 0.029 & 0.950 & - & - & -\\
\hline
y2sigma..Intercept. & 0.16 & 0.160 & 0.020 & 0.950 & 0.208 & 0.015 & 0.000\\
\hline
y2sigma.visit\_times & 0.01 & 0.010 & 0.005 & 0.965 & - & - & -\\
\hline
b\_sd4 & 0.16 & 0.158 & 0.018 & 0.965 & - & - & -\\
\hline
e.binary\_cov & 0.93 & 0.902 & 0.135 & 0.930 & 0.893 & 0.136 & 0.950\\
\hline
e.norm\_cov & -2.30 & -2.258 & 0.112 & 0.925 & -2.201 & 0.109 & 0.815\\
\hline
e.value.y1. & -2.24 & -2.133 & 0.152 & 0.945 & -2.089 & 0.166 & 0.700\\
\hline
e.value.y1sigma & 1.08 & 0.804 & 0.209 & 0.830 & - & - & -\\
\hline
e.value.y2. & 0.55 & 0.503 & 0.199 & 0.985 & 0.573 & 0.255 & 0.910\\
\hline
e.value.y2sigma & -0.12 & 0.295 & 0.373 & 0.995 & - & - & -\\
\end{comment}

\textbf{Longitudinal} - $y_1$ &&&&&&\\
$\beta_0^{\mu_1}$ - Intercept & 2.190 & 2.188 & 0.027 & 0.950 & 2.190 & 0.027 & 0.945\\
$\beta_1^{\mu_1}$ - Time & -0.040 & -0.040 & 0.002 & 0.950 & -0.040 & 0.002 & 0.960\\
$\tau^{\mu_1}$ - Random effect SD & 0.810 & 0.813 & 0.019 & 0.970 & 0.815 & 0.018 & 0.975\\
%\hline
$\beta_0^{\sigma_1}$ - Intercept & -1.360 & -1.366 & 0.028 & 0.965 & -1.238 & 0.029 & -\\
$\beta_1^{\sigma_1}$  - Time & -0.020 & -0.020 & 0.007 & 0.940 & - & - & -\\
$\tau^{\sigma_1}$ - Random effect SD& 0.440 & 0.441 & 0.020 & 0.950 & - & - & -\\
\hline
\textbf{Longitudinal - $y_2$} &&&&&&\\
$\beta_0^{\mu_2}$ - Intercept & 1.040 & 1.042 & 0.031 & 0.960 & 1.043 & 0.032 & 0.965\\
$\beta_1^{\mu_2}$  - Time & 0.350 & 0.350 & 0.009 & 0.910 & 0.351 & 0.009 & 0.930\\
$\tau^{\mu_2}$ - Random effect SD & 0.520 & 0.518 & 0.029 & 0.945 & 0.516 & 0.030 & 0.950\\
%\hline
$\beta_0^{\sigma_2}$ - Intercept & 0.160 & 0.160 & 0.020 & 0.965 & 0.208 & 0.014 & -\\
$\beta_1^{\sigma_2}$  - Time & 0.010 & 0.010 & 0.006 & 0.935 & - & - & -\\
$\tau^{\sigma_2}$ - Random effect SD & 0.160 & 0.159 & 0.020 & 0.965 & - & - & -\\
\hline
\textbf{Event} &&&&&&\\
$\gamma_1$ - Binary & 0.930 & 0.900 & 0.119 & 0.980 & 0.887 & 0.121 & 0.990\\
$\gamma_2$ - Normal & -2.300 & -2.275 & 0.106 & 0.980 & -2.215 & 0.103 & 0.855\\
%\hline
$\alpha^{\mu_1}$ - Mean of $y_1$ & -2.240 & -2.148 & 0.141 & 0.960 & -2.094 & 0.158 & 0.745\\
$\alpha^{\sigma_1}$ - WIV of $y_1$ & 1.080 & 0.828 & 0.206 & 0.860 & - & - & -\\
$\alpha^{\mu_2}$ - Mean of $y_2$ & 0.550 & 0.511 & 0.196 & 1.000 & 0.560 & 0.275 & 0.845\\
$\alpha^{\sigma_2}$ - WIV of $y_2$ & -0.120 & 0.243 & 0.399 & 0.995 & - & - & -\\
\hline
\end{tabular}
    \caption{Simulation results. For each model, the mean and standard devation (eSD) of the 200 posterior means are reported, alongside the coverage probability (proportion of times the input value is included in the 95\% credible interval).}
    \label{tab:simpar}
\end{table}

\section{Application}
\label{sec:appl}

Cystic fibrosis (CF) is a genetic condition due to a mutation in the cystic fibrosis transmembrane conductance regulator \textit{CFTR} gene (responsible for the movement of salt in cells), which results in mucus accumulation in the lungs and other organs. Infections and chronic inflammation due to CF often lead to a poor quality of life and lower life expectancy than the general population. Approximately 11,000 people in the UK are currently affected by cystic fibrosis \citep{cftrustreport2023}. Lung function is the most relevant biomarker in CF monitoring and its decline is associated with a higher mortality risk. Malnutrition is often observed in people with cystic fibrosis, as it is correlated with worse lung function and increased mortality \citep{nagy2022malnutrition}. In this work, we aim to provide some insights about the progression of CF, by jointly modelling WIV in lung function, malnutrition and their association with mortality.

\begin{figure}[hbtp]
    \centering
    \includegraphics[width=0.5\linewidth]{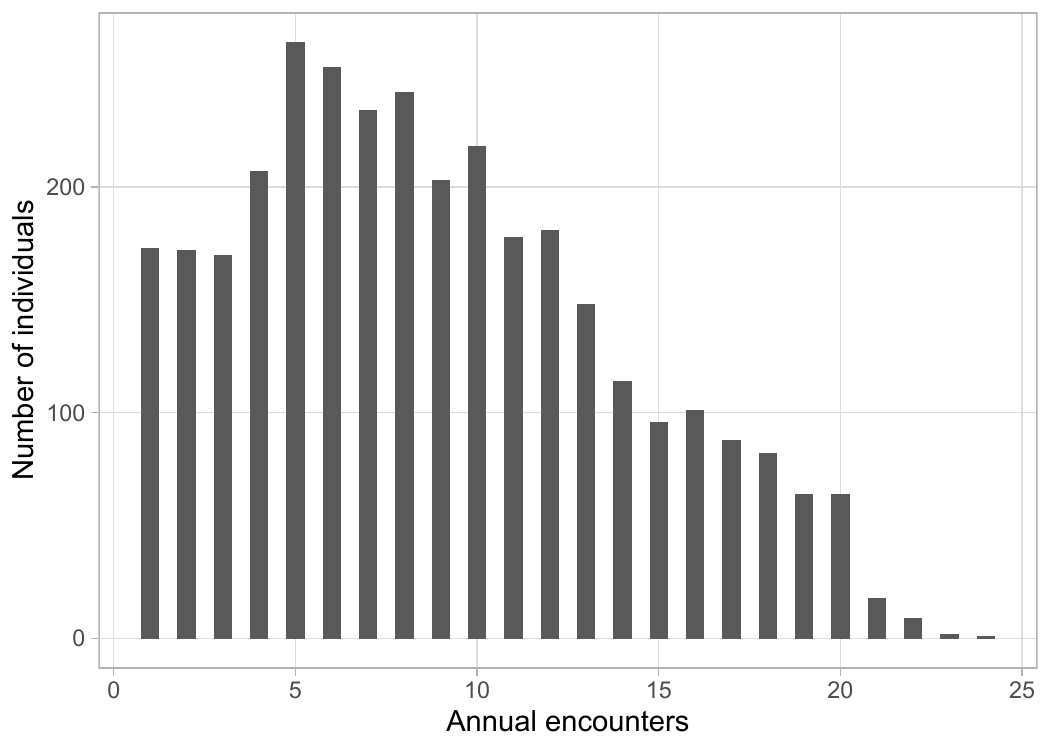}
    \caption{Histogram of annual encounters in the adult women subset.}
    \label{fig:hist_encounters}
\end{figure}

The data comes from the UK CF registry, containing anonymised electronic health records from people with CF between 1996 and 2020. Clinically stable individuals are called to a clinic for an annual encounter, during which a large range of variables about the lungs and other organs are recorded. We focus the analysis on women with at least one record in the registry during the follow-up period while they are between 18 and 49 years old (which accounts for almost the totality of the records) and we introduce administrative censoring at 50 years old (or at the end of the study period) if they have not experienced the event by then. The same inclusion criteria as in \citet{palma2024factors} were used in this study. The number of individuals in the dataset is 3282, for a total number of annual visits of 29094. The distribution of the number of encounters for each woman is shown in \Cref{fig:hist_encounters}: it ranges between 1 and 25 (with a median of 8 encounters and interquantile range $[5; 12]$), with a non-negligible share of women having very few encounters.

\begin{figure}[hbtp]
    \centering
    \includegraphics[width=0.5\linewidth]{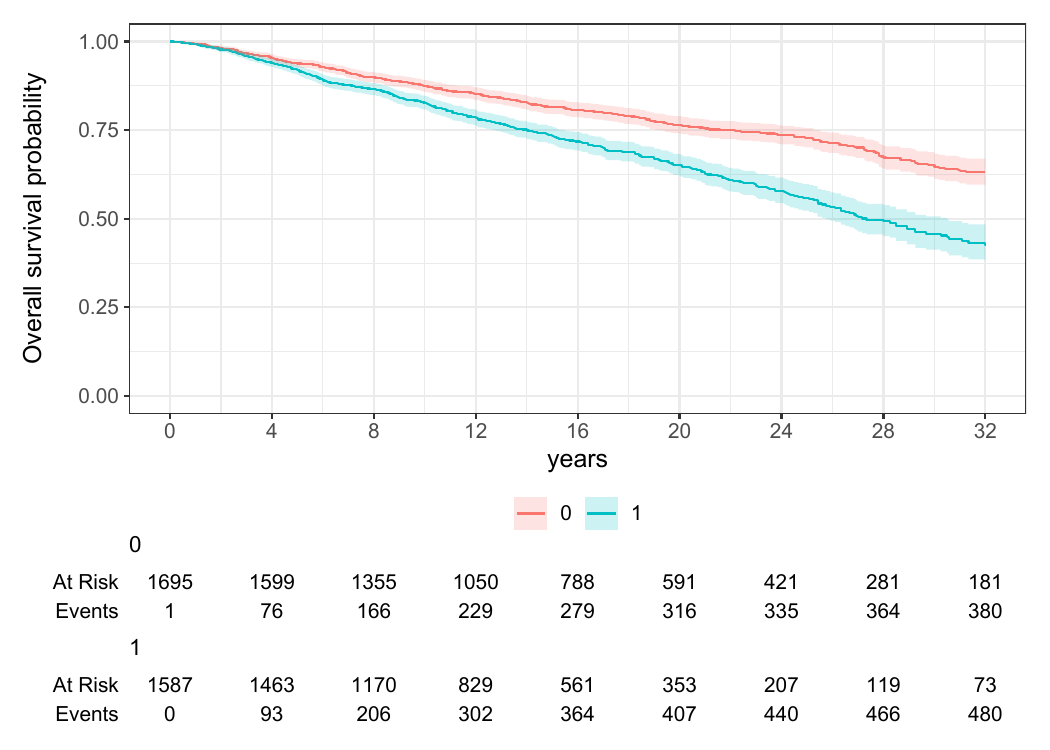}
    \caption{Kaplan-Meier curves by F508 (0: non homozygous; 1: homozygous).}
    \label{fig:KM_byf508}
\end{figure}

The F508 (often referred to as F508del) mutation in the CFTR gene is the most prevalent genetic alteration causing cystic fibrosis. We considered a binary variable for F508 (homozygous - 2 alleles vs.\ non-homozygous - 0,1, or missing information). The Kaplan-Meier curves by number of F508 alleles \Cref{fig:KM_byf508} show that those with two F508 alleles (F508 homozygous) women tend to have lower survival probability across almost all the time range. In further analysis (not shown here), no relevant differences between the Kaplan-Meier curves for the subgroups of non-homozygous were observed.

We build a bivariate joint model as in \Cref{eq:cfmodel} for the $i$-th individual, $j$-th encounter and $k$-th biomarker. We consider a measure of lung function (FEV\textsubscript{1}, forced expiratory volume in 1 second, measured in liters of air) as first biomarker ($k =1$) and body mass index (BMI) as second ($k=2$). Both the biomarkers (and their within-individual variability) are modelled as function of age and a random intercept, with baseline covariates as F508 class and age at diagnosis (before/after first year of life). In the time-to-event submodel, we include the current values for both means and variabilities for FEV\textsubscript{1} and BMI, as well as F508 class and age at diagnosis. 
We choose 6 cubic B-spline basis functions to flexibly model the log baseline hazard.
\begin{align}
\label{eq:cfmodel}
    %y_{ijk} &= \beta^{\mu_k}_0 + \beta^{\mu_k}_{1}\text{age}_{ij} + b^{\mu_k}_{0i} + \varepsilon_{ijk} \notag\\
    %\log(\sigma_{ijk}) &= \beta^{\sigma_k}_0 + \beta^{\sigma_k}_{1}\text{age}_{ij} + b^{\sigma_k}_{0i}\notag\\
    y_{ijk} &= \beta^{\mu_k}_0 + \beta^{\mu_k}_{1}\text{age}_{ij} +\beta^{\mu_k}_{2}\text{Diag1}_i + \beta^{\mu_k}_{3}\text{F508hom}_i + b^{\mu_k}_{0i} + \varepsilon_{ijk} \notag\\
    \log(\sigma_{ijk}) &= \beta^{\sigma_k}_0 + \beta^{\sigma_k}_{1}\text{age}_{ij} +  \beta^{\sigma_k}_{2}\text{Diag1}_i +\beta^{\sigma_k}_{3}\text{F508hom}_i + b^{\sigma_k}_{0i}\notag\\
    h_i(t) &= \exp\{\log h_0(t) + 
    \gamma_1\text{Diag1}_i +\gamma_2\text{F508hom}_i % +\gamma_3\text{YearOfBirth}_i    %+ \notag\\ & 
    + \alpha_1\mu_{i1}(t) + \alpha_2\sigma_{i1}(t) +\alpha_3\mu_{i2}(t) +\alpha_4\sigma_{i2}(t)\}
\end{align}
We fit the model using Stan, with 2 chains, 1000 burn-in iterations and a total of 2000 iterations. The posterior estimates and 95\% credible intervals of all variables in the longitudinal submodel are reported in \Cref{tab:FULLlong}.

\begin{table}[hbtp]
    \centering
\begin{tabular}{l|r|r|r|rr|r}
\hline
  & Mean & MCMC-SE & SD & 2.5\% & 97.5\% & $\hat{R}$\\
\hline
\textbf{Longitudinal - FEV\textsubscript{1} mean} &&&&&&\\
$\beta_0^{\mu_1}$ - Intercept & 2.172 & 0.004 & 0.026 & 2.121 & 2.223 & 1.025\\
$\beta_1^{\mu_1}$ - Age & -0.040 & 0.000 & 0.000 & -0.041 & -0.040 & 1.001\\
$\beta_2^{\mu_1}$ - Diagnosed after 1 year old & 0.326 & 0.004 & 0.030 & 0.272 & 0.384 & 1.004\\
$\beta_3^{\mu_1}$ - F508 homozygous & -0.250 & 0.004 & 0.026 & -0.300 & -0.197 & 1.033\\
$\tau^{\mu_1}$ - Random effect SD & 0.792 & 0.001 & 0.009 & 0.774 & 0.809 & 1.037\\
\textbf{Longitudinal - FEV\textsubscript{1} log(SD)} &&&&&&\\
$\beta_0^{\sigma_1}$ - Intercept & -1.334 & 0.001 & 0.019 & -1.371 & -1.295 & 1.000\\
$\beta_1^{\sigma_1}$ - Age & -0.018 & 0.000 & 0.001 & -0.020 & -0.016 & 1.000\\
$\beta_2^{\sigma_1}$ - Diagnosed after 1 year old & 0.001 & 0.001 & 0.021 & -0.039 & 0.041 & 1.002\\
$\beta_3^{\sigma_1}$ - F508 homozygous & 0.037 & 0.001 & 0.020 & -0.002 & 0.077 & 1.000\\
$\tau^{\sigma_1}$ - Random effect SD & 0.436 & 0.000 & 0.008 & 0.420 & 0.453 & 1.007\\
\hline
\textbf{Longitudinal - BMI mean} &&&&&&\\
$\beta_0^{\mu_2}$ - Intercept & 21.182 & 0.012 & 0.104 & 20.982 & 21.387 & 1.013\\
$\beta_1^{\mu_2}$ - Age & 0.045 & 0.000 & 0.002 & 0.041 & 0.049 & 0.999\\
$\beta_2^{\mu_2}$ - Diagnosed after 1 year old  & 0.898 & 0.015 & 0.119 & 0.667 & 1.138 & 1.026\\
$\beta_3^{\mu_2}$ - F508 homozygous & -0.972 & 0.014 & 0.112 & -1.199 & -0.760 & 1.002\\
$\tau^{\mu_2}$ - Random effect SD & 3.315 & 0.003 & 0.046 & 3.229 & 3.409 & 1.004\\
\textbf{Longitudinal - BMI log(SD)} &&&&&&\\
$\beta_0^{\sigma_2}$ - Intercept & 0.237 & 0.001 & 0.020 & 0.197 & 0.276 & 1.004\\
$\beta_1^{\sigma_2}$ - Age & 0.001 & 0.000 & 0.001 & -0.001 & 0.003 & 1.000\\
$\beta_2^{\sigma_2}$ - Diagnosed after 1 year old  & 0.010 & 0.001 & 0.020 & -0.029 & 0.049 & 1.004\\
$\beta_3^{\sigma_2}$ - F508 homozygous & -0.078 & 0.001 & 0.020 & -0.117 & -0.041 & 1.006\\
$\tau^{\sigma_2}$ - Random effect SD & 0.458 & 0.000 & 0.008 & 0.442 & 0.474 & 1.000\\
\hline
%\textbf{Event - Baseline covariates} &&&&&&\\
%$w_1$ - Diagnosed after 1 year old & -0.337 & 0.001 & 0.079 & -0.493 & -0.186 & 0.999\\
%$w_2$ - F508 homozygous & 0.057 & 0.001 & 0.073 & -0.090 & 0.201 & 1.001\\
%\hline
%\textbf{Association parameters (log hazard scale)} &&&&&&\\
%$\alpha^{\mu_1}$ - Mean FEV\textsubscript{1} (in L) & -2.673 & 0.003 & 0.115 & -2.901 & -2.454 & 1.004\\
%$\alpha^{\sigma_1}$ - WIV of FEV\textsubscript{1} (in L) & 3.178 & 0.014 & 0.427 & 2.374 & 4.045 & 1.003\\
%$\alpha^{\mu_2}$ - Mean BMI & -0.089 & 0.001 & 0.023 & -0.134 & -0.043 & 1.010\\
%$\alpha^{\sigma_2}$ - WIV of BMI & 0.179 & 0.004 & 0.097 & -0.013 & 0.364 & 1.005\\
%\hline
\end{tabular}
    \caption{Results for the CF longitudinal submodel: posterior mean and standard deviation, MCMC standard error (defined as the ratio between standard deviation and square root of effective sample size), 95\% credible interval and $\hat{R}$ statistic.}\label{tab:FULLlong}
\end{table}

\begin{figure}[hbtp]
    \centering
    \includegraphics[width=0.49\linewidth]{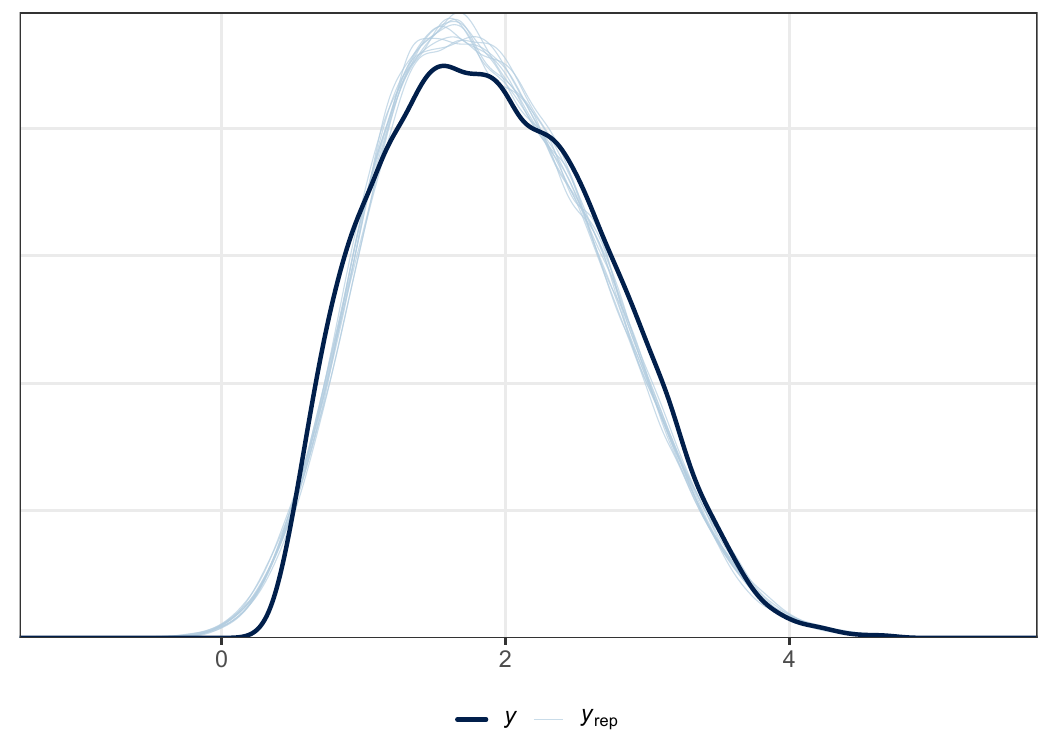}\hfill
    \includegraphics[width=0.49\linewidth]{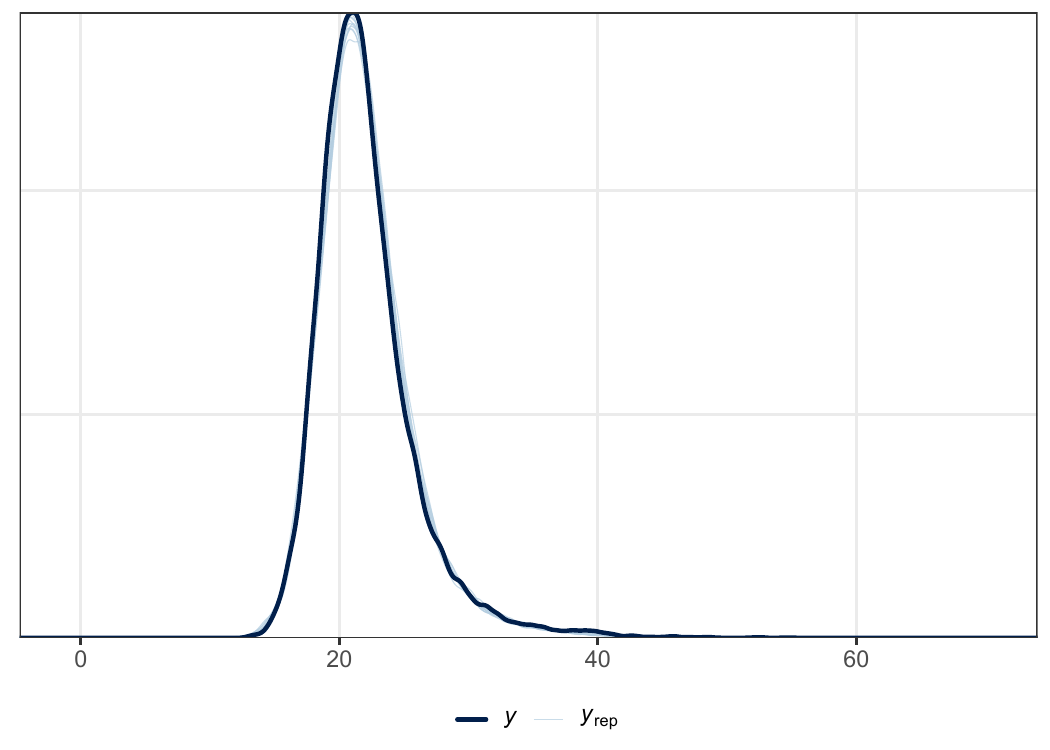}
    \caption{Posterior predictive checks for longitudinal submodel (left: FEV\textsubscript{1}; right: BMI).}
    \label{fig:ppchecks}
\end{figure}

As expected, FEV\textsubscript{1} tends to decrease at older ages; women with a later diagnosis show better lung function on average compared to those diagnosed soon after birth, and the same conclusion applies for non-homozygous women compared to homozygous ones. FEV\textsubscript{1} variability instead tends to decline as age increases (as shown also in \citealp{palma2024factors}) and no effect is found for the other covariates. For BMI, the mean tends to be increasing as age increases. Women with later diagnosis and F508 non-homozygous show on average higher BMI measurements than the other groups. In the variability submodel for BMI, no age change is estimated, while only F508 has a non-zero estimate, with lower WIV observed for homozygous women on average. The 95\% credible intervals of the four random effects are all far from zero, indicating that there is still additional variability in the mean and standard deviation submodels which is not captured by the covariates in the model. In addition, the posterior means of the correlation structure (reported in the Supplementary Material) for the random effects suggest that there is a positive correlation between the FEV\textsubscript{1} and BMI mean (approximately equal to 0.5), as well as for mean and WIV for both biomarkers. These results agree with the physiology of the problem: especially for FEV\textsubscript{1}, when an individual have lower functionality due to lung damage the fluctuations will also have a limited range. The longitudinal submodel fits the data well, as shown by the posterior predictive checks in \Cref{fig:ppchecks}, for both biomarkers. 

The estimates for the hazard ratios of the longitudinal biomarkers in the time-to-event submodels are reported in \Cref{tab:FULL_assoc}. The hazard ratios with FEV\textsubscript{1} are expressed in dL for better interpretability. 

\begin{comment}
\begin{table}[h!]\label{tab:SIMPLElong_assoc}
    \centering
\begin{tabular}{l|r|r|r}
\hline
  & HR & 2.5\% & 97.5\%\\
\hline
%$\alpha^{\mu_1}$ - 
Association with mean FEV\textsubscript{1} (in dL) & 0.767 & 0.749 & 0.783\\
\hline
%$\alpha^{\sigma_1}$ - 
Association with WIV of FEV\textsubscript{1} (in dL)& 1.363 & 1.253 & 1.477\\
\hline
%$\alpha^{\mu_2}$ - 
Association with mean BMI & 0.915 & 0.875 & 0.954\\
\hline
%$\alpha^{\sigma_2}$ - 
Association with WIV of BMI &  1.203 & 0.987 & 1.459\\
\hline
\end{tabular}
    \caption{Association between current values for longitudinal submodel and mortality}
\end{table}
\end{comment}

\begin{table}[h!]
    \centering
\begin{tabular}{l|r|rr}
  & HR & 2.5\% & 97.5\%\\
\hline
%$\alpha^{\mu_1}$ - 
Association with mean FEV\textsubscript{1} (in dL) & 0.765 & 0.748 & 0.782\\
%\hline
%$\alpha^{\sigma_1}$ - 
Association with WIV of FEV\textsubscript{1} (in dL)&  1.374 & 1.268 & 1.499\\
%\hline
%$\alpha^{\mu_2}$ - 
Association with mean BMI & 0.915 & 0.875 & 0.958\\
%\hline
%$\alpha^{\sigma_2}$ - 
Association with WIV of BMI & 1.196 & 0.987 & 1.439\\
\hline
\end{tabular}
    \caption{Association between current values for longitudinal biomarkers and mortality.}\label{tab:FULL_assoc}
\end{table}

\begin{figure}[hbtp]
    \centering
    \includegraphics[width=0.8\linewidth]{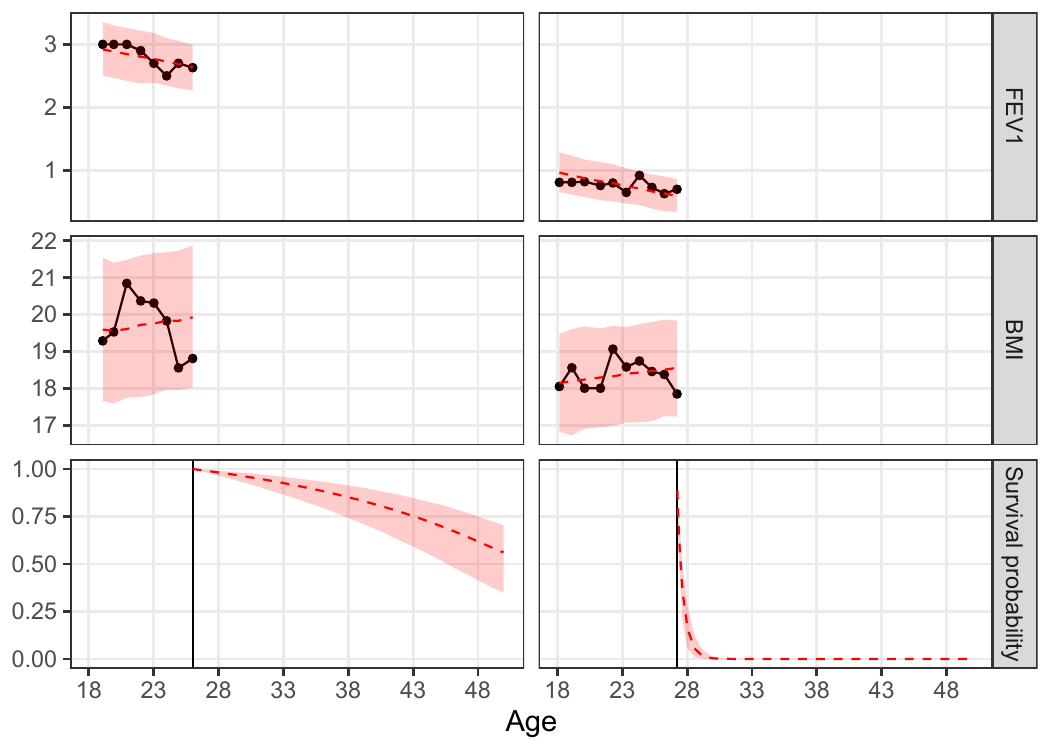}
    \caption{Predictions for two random individuals in the dataset (left: censored; right: dead). Black line and dots are the observed data, while the red line is the model prediction and the red bands show the variability.}
    \label{fig:longpred}
\end{figure}

Both FEV\textsubscript{1} and BMI mean show a hazard ratio (HR) lower than 1, indicating that worse lung functionality and lower body mass index are both associated with higher risk of death. In addition, the within-individual variability for FEV\textsubscript{1} shows a 95\% credible interval above 1, indicating also higher FEV\textsubscript{1} variability is associated with increased mortality. No clear evidence of association with BMI variability was found. 
We also display the fitted values for two randomly selected individuals (with a similar age at last observation) in the dataset in \Cref{fig:longpred}. The figure clearly shows the WIV difference (especially in BMI) between the two individuals. The predicted survival curves are also very different: %one individual shows a better lung functionality and nutrition status than the other, and consequently 
the woman with better lung functionality and nutrition status shows a higher survival probability than the other woman across all the age range.

\section{Discussion}
\label{sec:discussion}

%Within-individual variability has received attention recently in longitudinal data analysis due to its potential association with time-to-event outcomes, in clinical applications as well as in statistical methodology\todo{Remove sentence as it is a repetition}. 
In this work, we proposed a joint model which allows the quantification of within-individual variability in multivariate longitudinal data and its association with a time-to-event  outcome. The proportional hazard model for the outcome accounts for the association with the means and the within-individual variabilities of the longitudinal biomarkers and the baseline hazard can be flexibly specified using B-splines. Our simulation study shows that when there is an association between the within-individual variability and the event, neglecting the corresponding coefficients leads to bias and loss of coverage of the other coefficients in the time-to-event submodel. In the cystic fibrosis application, we have shown that higher lung function within-individual variability (but not BMI variability) is associated with a higher risk of death, confirming preliminary work in \citet{palma2024eps6}.

One advantage of this model is the additive structure for each longitudinal biomarker, i.e.\ the ability to accommodate changes over time, baseline covariates and random effects as in a generalised additive model. Due to the shared random effect structure, one can easily model the individual means and variabilities, as well as correlations between subject-specific random effects. In addition, compared to the summary statistics often used in clinical settings, the use of random effects as subject-specific deviations from the population variability retains a similar level of interpretability, as one could e.g.\ read a positive random effect in the variability submodel as a higher WIV than the population standard deviation, accounting at the same time for time and covariate effects. This Bayesian joint model makes better use of all information available in the dataset, as all observations contribute to the estimation of the population standard deviation, without the need to exclude individuals with few visits and introduce immortal time bias.

There are some limitations in the modelling strategy proposed here. As pointed out in \citet{wang2024modeling} about location-scale models like the model proposed here and the one in \citet{barrett2019estimating}, the within-individual variability is no longer distinguishable from the measurement error. The quantification of variability proposed in \citealp{wang2024modeling} (based on the second derivative of the mean function) can nevertheless be integrated in the location-scale model to characterise different aspects of variability. 

Furthermore, while in theory the model can be extended to any number of longitudinal biomarkers, the estimation of the model might become difficult for a large number of random effects (as in most multivariate models with shared random effects). In our application we used only one random intercept for each parameter of the distribution, for a total of two random effects per biomarker. If one wants to include additional random effects over time (either as random slope or random effects for spline basis functions), or increase the number of biomarkers, the number of parameters in the model becomes large (as well as the parameters in the correlation matrix, which is specified to be unconstrained). We recommend to use a moderate number of biomarkers and, wherever possible, a parsimonious number of random effects (or constraints on the correlation structure). Alternatively, a more parsimonious specification like in \citealp{volkmann2023flexible} could help to reduce the computational burden. The current version of the code (available on \url{https://github.com/marcopalma3/JM-WIV}) can be also extended to include more flexibility, by including more than two outcomes, increase the number of random effects and introduce constraints on their correlation matrix. 

The joint model structure presented here might also be developed in other directions. For the longitudinal submodel, the assumption of normal distribution can be relaxed, drawing from the work on generalised additive models for location, scale and shape \citep{rigby2005GAMLSS}. Nevertheless, the direct interpretation of the standard deviation parameter as within-individual variability is not preserved. The definition of within-individual variability in non-normal settings (for example, in discrete distributions) is still an open question which needs further investigation, given also the large availability of discrete biomarkers in biomedical applications (see \citealp{parker2024modeling} for an example using a beta-binomial model).

For the cystic fibrosis application, by comparing the results obtained with this model and those based on summary statistics, one could confirm the relevance of within-individual variability in defining the hazard of an event, in a similar way as proposed in \citet{decourson2021impact}. In addition, the model could be further extended to include a competing risk approach for the analysis of transplant information. Large interest is also in the effect of recently introduced medications (triple combination therapies, with promising results) on the within-individual variability, by using a change-point structure in the variability submodel. In this direction, the availability of CF registries with different structures (annual measurements in UK, monthly records in Denmark, irregular encounters in the USA) might require an adaptation of the model to efficiently model WIV and the relationship with potentially informative measurement schedule. In addition, the results of the analysis presented here open avenues for a better characterisation of within-individual variability in disease progression and dynamic risk prediction to assess the effective contribution of WIV in this setting. In this framework, new approaches for efficient sequential WIV quantification in online data streaming (especially in electronic health data) need to be developed, in order to potentially introduce the use of these WIV approaches in clinical settings.

%\end{linenumbers}

\newpage
%\vspace{10cm}

\section*{Author contributions (CRediT)} 

Conceptualisation: all authors (equal contribution). 
Formal Analysis: MP. 
Funding acquisition: JKB, GM-T (lead contribution), RHK, AMW (supporting contribution). 
Methodology: MP, JKB, GM-T, RHK, AMW (equal contribution). 
Software: MP. 
Supervision: JKB. 
Visualization: MP. 
Writing – original draft: MP. % (lead contribution), JKB (supporting contribution). 
Writing – review \& editing: all authors (equal contribution). 

%\vspace{-1.5em}
\section*{Data and code availability} 

This work used anonymised data from the UK Cystic Fibrosis Registry, which has research ethics approval (Research Ethics Committee reference number 07/Q0104/2). Use of the data was approved by the Registry Research Committee (data request 426). Data are available following application to the Registry Research Committee 
(www.cysticfibrosis.org.uk/the-work-we-do/uk-cf-registry/apply-for-data-from-the-uk-cf-registry). The R code is available at %\url{https://github.com/marcopalma3/rstanjmwiv} and 
\url{https://github.com/marcopalma3/JM-WIV}.

%\vspace{-1.5em}
\section*{Source of funding} 

MP was supported by the UK Medical Research Council (MRC) grant “Looking beyond the mean: what within-person variability can tell us about dementia, cardiovascular disease and cystic fibrosis” (MR/V020595/1). JKB was supported through the UK Medical Research Council programme (grant MC\_UU\_00002/5) and Unit theme (grant MC\_UU\_00040/02 – Precision Medicine) funding. RHK was funded by UK Research and Innovation (Future Leaders Fellowship MR/X015017/1). RS was supported by the National Heart, Lung and Blood Institute of the National Institutes of Health (R01 HL141286) and Cystic Fibrosis Foundation (Naren19R0 and SZCZES22AB0). GM-T acknowledges the support of the Osteopathic Heritage Foundation through funding for the Osteopathic Heritage Foundation Ralph S. Licklider, D.O., Research Endowment in the Heritage College of Osteopathic Medicine. 

For the purpose of open access, the authors have applied a Creative Commons Attribution (CC BY) licence to any Author Accepted Manuscript version arising from this submission. 

%\vspace{-1.5em}
\section*{Acknowledgements} 

We thank the Cystic Fibrosis Epidemiological Network (CF-EpiNet) Strategic Research Centre data group and Elaine Gunn for contributions to data preparation and cleaning. %We thank the reviewers for their detailed and insightful feedback.  

%\vspace{-1.5em}
\section*{Competing interests}
 
JKB has received research funding for unrelated work from F. Hoffmann-La Roche Ltd.

%\vspace{-1.5em}
%\section*{Publication history}
 
%An early version of this manuscript is available in medRxiv: doi: https://doi.org/10.1101/2023.05.12.23289768.

\bigskip
\newpage

\bibliographystyle{agsm}
\bibliography{JM_WIV}

\newpage
\section*{\centering Supplementary Material}
\setcounter{table}{0}

\renewcommand{\thefigure}{S.\arabic{figure}}
\renewcommand{\thetable}{S.\arabic{table}}

%\title{Supplementary Material\\ \normalsize A Bayesian location-scale joint model for time-to-event and multivariate longitudinal data with association based on within-individual variability}
%\author{by Palma et al. (2025)}
%\date{}

\maketitle

\section*{Simulation study}
The matrix below shows the correlation between random effects used to generate the data: first two rows/columns refer to mean and variability of the first biomarker.

\begin{center}
%$\mathcal{P}$ = 
$\begin{bmatrix}
1  & 0.129 & 0.502 & 0.019\\
0.129 &  1 & -0.006 & 0.243\\
0.502 & -0.006 & 1 & 0.491\\
0.019 & 0.243 & 0.491 & 1\\ 
    \end{bmatrix}$
\end{center}

\section*{Additional results}

\begin{comment}
    
\begin{table}[hbtp]
    \centering
\begin{tabular}{l|r|rr}
  & Mean & 2.5\% & 97.5\%\\
\hline
$\tau^{\mu_1}$ - Mean FEV\textsubscript{1}
& 0.792 & 0.774 & 0.809\\
$\tau^{\sigma_1}$ - WIV of FEV\textsubscript{1}
 & 0.436 & 0.420 & 0.453\\
$\tau^{\mu_2}$ - Mean BMI & 3.315 & 3.229 & 3.409\\
$\tau^{\sigma_2}$ - WIV of BMI & 0.458 & 0.442 & 0.474\\
\hline
\end{tabular}
    \caption{Posterior estimates and 95\% credible intervals for standard deviation of random intercepts for location and scale submodel for both longitudinal biomarkers.}
    \label{tab:ranefs}
\end{table}  
\end{comment}

The matrix below shows the correlation between random effects: first two rows/columns refer to FEV\textsubscript{1} mean and variability.

\begin{center}
%$\mathcal{P}$ = 
$\begin{bmatrix}
1 & 0.147 & 0.475 & 0.005\\
0.147 & 1 & 0.002 & 0.232\\
0.475 & 0.002 & 1 & 0.494\\
0.005 & 0.232 & 0.494 & 1\\
    \end{bmatrix}$
\end{center}

\begin{table}[hbtp]\label{tab:FULL}
    \centering
\begin{tabular}{l|r|r|r|rr|r}
\hline
  & Mean & MCMC-SE & SD & 2.5\% & 97.5\% & $\hat{R}$\\
\hline
\textbf{Longitudinal - FEV\textsubscript{1} mean} &&&&&&\\
$\beta_0^{\mu_1}$ - Intercept & 2.172 & 0.004 & 0.026 & 2.121 & 2.223 & 1.025\\
$\beta_1^{\mu_1}$ - Age & -0.040 & 0.000 & 0.000 & -0.041 & -0.040 & 1.001\\
$\beta_2^{\mu_1}$ - Diagnosed after 1 year old & 0.326 & 0.004 & 0.030 & 0.272 & 0.384 & 1.004\\
$\beta_3^{\mu_1}$ - F508 homozygous & -0.250 & 0.004 & 0.026 & -0.300 & -0.197 & 1.033\\
$\tau^{\mu_1}$ - Random effect SD & 0.792 & 0.001 & 0.009 & 0.774 & 0.809 & 1.037\\
\textbf{Longitudinal - FEV\textsubscript{1} log(SD)} &&&&&&\\
$\beta_0^{\sigma_1}$ - Intercept & -1.334 & 0.001 & 0.019 & -1.371 & -1.295 & 1.000\\
$\beta_1^{\sigma_1}$ - Age & -0.018 & 0.000 & 0.001 & -0.020 & -0.016 & 1.000\\
$\beta_2^{\sigma_1}$ - Diagnosed after 1 year old & 0.001 & 0.001 & 0.021 & -0.039 & 0.041 & 1.002\\
$\beta_3^{\sigma_1}$ - F508 homozygous & 0.037 & 0.001 & 0.020 & -0.002 & 0.077 & 1.000\\
$\tau^{\sigma_1}$ - Random effect SD & 0.436 & 0.000 & 0.008 & 0.420 & 0.453 & 1.007\\
\hline
\textbf{Longitudinal - BMI mean} &&&&&&\\
$\beta_0^{\mu_2}$ - Intercept & 21.182 & 0.012 & 0.104 & 20.982 & 21.387 & 1.013\\
$\beta_1^{\mu_2}$ - Age & 0.045 & 0.000 & 0.002 & 0.041 & 0.049 & 0.999\\
$\beta_2^{\mu_2}$ - Diagnosed after 1 year old  & 0.898 & 0.015 & 0.119 & 0.667 & 1.138 & 1.026\\
$\beta_3^{\mu_2}$ - F508 homozygous & -0.972 & 0.014 & 0.112 & -1.199 & -0.760 & 1.002\\
$\tau^{\mu_2}$ - Random effect SD & 3.315 & 0.003 & 0.046 & 3.229 & 3.409 & 1.004\\
\textbf{Longitudinal - BMI log(SD)} &&&&&&\\
$\beta_0^{\sigma_2}$ - Intercept & 0.237 & 0.001 & 0.020 & 0.197 & 0.276 & 1.004\\
$\beta_1^{\sigma_2}$ - Age & 0.001 & 0.000 & 0.001 & -0.001 & 0.003 & 1.000\\
$\beta_2^{\sigma_2}$ - Diagnosed after 1 year old  & 0.010 & 0.001 & 0.020 & -0.029 & 0.049 & 1.004\\
$\beta_3^{\sigma_2}$ - F508 homozygous & -0.078 & 0.001 & 0.020 & -0.117 & -0.041 & 1.006\\
$\tau^{\sigma_2}$ - Random effect SD & 0.458 & 0.000 & 0.008 & 0.442 & 0.474 & 1.000\\
\hline
\textbf{Event - Baseline covariates} &&&&&&\\
$\gamma_1$ - Diagnosed after 1 year old & -0.337 & 0.001 & 0.079 & -0.493 & -0.186 & 0.999\\
$\gamma_2$ - F508 homozygous & 0.057 & 0.001 & 0.073 & -0.090 & 0.201 & 1.001\\
\hline
\textbf{Association parameters (log hazard scale)} &&&&&&\\
$\alpha^{\mu_1}$ - Mean FEV\textsubscript{1} (in L) & -2.673 & 0.003 & 0.115 & -2.901 & -2.454 & 1.004\\
$\alpha^{\sigma_1}$ - WIV of FEV\textsubscript{1} (in L) & 3.178 & 0.014 & 0.427 & 2.374 & 4.045 & 1.003\\
$\alpha^{\mu_2}$ - Mean BMI & -0.089 & 0.001 & 0.023 & -0.134 & -0.043 & 1.010\\
$\alpha^{\sigma_2}$ - WIV of BMI & 0.179 & 0.004 & 0.097 & -0.013 & 0.364 & 1.005\\
\hline
\end{tabular}
    \caption{Results for the CF model: posterior mean and standard deviation, MCMC standard error (defined as the ratio between standard deviation and square root of effective sample size), 95\% credible interval and $\hat{R}$ statistic.}
\end{table}

\end{document}